\pdfoutput=1
\documentclass{aa}

\usepackage[varg]{txfonts}
\usepackage{lineno}
\usepackage{natbib}
\usepackage{graphicx}
\usepackage{amsmath}
\usepackage{amssymb}
\usepackage{hyperref}
\hypersetup{
    colorlinks   = true,
    urlcolor     = blue,
    linkcolor    = blue,
    citecolor   = blue
}
\usepackage{multirow}
\usepackage{tabularx}
\usepackage{bm}
\usepackage{xcolor}
\usepackage[export]{adjustbox}
\usepackage{url}

\usepackage{tikz}
\usepackage{microtype}

\begin{document}

   \title{Probabilistic redshift estimation of undetected galaxies from multi-band background light maps}

   \author{Shun-Sheng Li\inst{1,2,3}
   \and Hendrik Hildebrandt\inst{1}
   \and Ludovic Van Waerbeke\inst{4}}
   \institute{Ruhr University Bochum, Faculty of Physics and Astronomy, Astronomical Institute (AIRUB), German Centre for Cosmological Lensing, 44780 Bochum, Germany\\
   \email{liss@stanford.edu}
   \and
   Kavli Institute for Particle Astrophysics and Cosmology, Stanford University, Stanford, CA 94305, USA
   \and
   SLAC National Accelerator Laboratory, Menlo Park, CA 94025, USA
   \and
   Department of Physics and Astronomy, The University of British Columbia, 6224 Agricultural Road, V6T 1Z1, Vancouver, Canada
   }

   \date{Accepted 23 October 2026}

  \abstract
  {Accurate knowledge of the redshift distributions of undetected galaxy populations is essential for extracting the cosmological information encoded in the cosmic optical background. We present the first framework for estimating these distributions directly from multi-band maps of unresolved background light, using conditional normalising flows trained on realistic image simulations. For imaging quality comparable to that expected from the \textit{Euclid} survey and the Vera C. Rubin Observatory Legacy Survey of Space and Time (LSST), our model achieves sub-per cent accuracy in both the mean and standard deviation of the flux-weighted redshift distributions when the training and target samples are statistically well matched. Furthermore, the network proves highly resilient to incomplete photometric coverage when combined with tailored data imputation strategies. To address the challenge of statistical mismatches between simulations and real observations, we propose incorporating observational quantities into the target variables, providing a built-in diagnostic to assess model reliability and uncertainty from the predictions themselves. We also demonstrate the feasibility of map-based tomographic analyses, showing that the model retains sufficient pixel-level detail to recover redshift distributions for tomographically defined subsamples with near sub-per cent accuracy across all bins. }


   \maketitle
\nolinenumbers

\section{Introduction}

Cosmic electromagnetic background radiation encodes rich information about the Universe and has long been recognised as a powerful probe of both the early Universe and large-scale structure (see \citealt{Cooray2016RSOS....350555C,Hill2018ApSpe..72..663H} for some reviews). Besides the well-established field of the cosmic microwave background (CMB), which represents a milestone in modern precision cosmology~(e.g.~\citealt{Bennett2013ApJS..208...20B,Planck2020AA...641A...1P}), studies of diffuse backgrounds at other wavelengths are also advancing rapidly. In particular, the cosmic infrared background (CIB; $1{\mu\rm m}\lesssim\lambda\lesssim500{\mu\rm m}$) has emerged as an active field of study (see \citealt{Hauser2001ARAA..39..249H,Lagache2005ARAA..43..727L,Kashlinsky2005PhR...409..361K} for some reviews). Although not as mature as CMB studies, observations and analyses of the CIB have progressed steadily through several generations of space-based instruments~(e.g.~\citealt{Puget1996AA...308L...5P,Hauser1998ApJ...508...25H,Genzel2000ARAA..38..761G,Dole2006AA...451..417D,Matsuura2011ApJ...737....2M,Matsuura2017ApJ...839....7M,Carleton2022AJ....164..170C}).

Studies of the cosmic optical background (COB; $0.4{\mu\rm m}\lesssim\lambda\lesssim1{\mu\rm m}$) have been hindered primarily by bright foreground contamination (e.g. the zodiacal light), further exacerbated by the Earth’s atmosphere (e.g.~\citealt{Windhorst2022AJ....164..141W}). This situation is now changing with rapidly improving space-based observations, including the Hubble Space Telescope~\citep{Tompkins2026MNRAS.tmp...42T}, the James Webb Space Telescope~\citep{Windhorst2023AJ....165...13W}, the \textit{Euclid} mission~\citep{Euclid2024AA...689A.294E}, the Spectro-Photometer for the History of the Universe, Epoch of Reionization and Ices Explorer (SPHEREx) survey~\citep{Dor2014arXiv1412.4872D}, the Long-range Reconnaissance Imager on the New Horizons spacecraft~\citep{Postman2024ApJ...972...95P}, and the planned \textit{Nancy Grace Roman} Space Telescope~\citep{Akeson2019arXiv190205569A}. These instrumental advances are complemented by rapid progress in image-processing techniques, driven in large part by low-surface-brightness science and the stability of space-based photometry (e.g.~\citealt{Euclid2022AA...662A.112E,Euclid2022AA...657A..92E,Liu2023ApJ...953....7L,Cuillandre2025AA...697A...6C}).

Robust measurements of the COB would open a new observational window for cosmology, complementary to conventional galaxy surveys that rely on detected sources. This is particularly compelling given that the vast majority of galaxies are expected to remain undetected even in Stage-IV surveys such as \textit{Euclid} and the Vera C. Rubin Observatory Legacy Survey of Space and Time (LSST)~\citep{Conselice2016ApJ...830...83C}. The diffuse COB is therefore expected to contain a wealth of information about the undetected galaxy population and the integrated history of cosmic light production. Beyond serving as another tracer of large-scale structure through the integrated light of undetected faint galaxies, the COB is also expected to carry information about the first stars and galaxies during the epoch of reionisation (e.g.~\citealt{Kashlinsky2005Natur.438...45K,Mitchell2015NatCo...6.7945M}), and to provide an avenue for testing certain dark matter candidates predicted to generate faint, diffuse electromagnetic emission peaking at optical wavelengths~(e.g.~ \citealt{Gong2016ApJ...825..104G,Kalashev2019PhRvD..99b3002K,Caputo2021JCAP...05..046C,Carenza2023PhRvD.107h3032C,Majidi2024JCAP...09..045M}).

To achieve these scientific goals, signals originating from different sources must be effectively separated. A key approach is to leverage statistical methods based on intensity maps to enhance detection significance (e.g.\ \citealt{Kashlinsky2005Natur.438...45K,Kashlinsky2012ApJ...753...63K,Matsumoto2011ApJ...742..124M}). Accurate estimation of the redshift distributions of sources contributing to the integrated signals in intensity maps would improve this process, facilitating component separation through correlation analyses, aiding signal interpretation, and ultimately enabling studies of galaxy formation and cosmological inference (e.g.~\citealt{Lim2023MNRAS.525.1443L,Euclid2026arXiv260121111E}).

In this paper, we address this challenge by exploring methods for inferring the redshift distributions of undetected galaxy populations directly from multi-band maps of unresolved background light. We focus on the synergy between the \textit{Euclid} mission and LSST, which together provide ten-band photometry spanning optical to near-infrared wavelengths. As a space-based mission, \textit{Euclid} offers a superior point spread function (PSF) and a stable, atmosphere-free noise background, making it well suited for studying the low-surface-brightness Universe~\citep{Euclid2022AA...662A.112E,Euclid2022AA...657A..92E}. The LSST complements this with deep, wide-field ground-based imaging in six optical bands, providing the broad-band colour information essential for photometric redshift estimation \citep{Ivezic2019ApJ...873..111I}. Building on insights from the well-established photometric redshift framework developed for detected galaxies, and leveraging modern image simulations and machine learning techniques, we demonstrate the feasibility of recovering probabilistic redshift distributions for undetected samples under the observational noise levels expected for \textit{Euclid} and LSST. To the best of our knowledge, this work presents the first attempt to estimate redshift distributions of undetected galaxies directly from maps of unresolved background light. It constitutes an essential step towards future cosmological analyses that exploit undetected galaxy populations as tracers.

The remainder of this paper is structured as follows. Section~\ref{Sec:Sim} describes the construction of multi-band mock background light maps with noise properties designed to mimic those expected for \textit{Euclid} and LSST. Section~\ref{Sec:method} introduces our redshift estimation framework based on conditional normalising flows. Section~\ref{Sec:res} presents the results, including quantitative assessments of model performance for both overall redshift distributions and potential tomographic binning schemes, as well as methods for quantifying prediction uncertainties. We conclude in Section~\ref{Sec:con} with a discussion of the implications of our findings and the further developments needed for practical application.

\section{Simulations}
\label{Sec:Sim}

  \begin{figure*}
  \centering
  \includegraphics[width=0.90\textwidth]{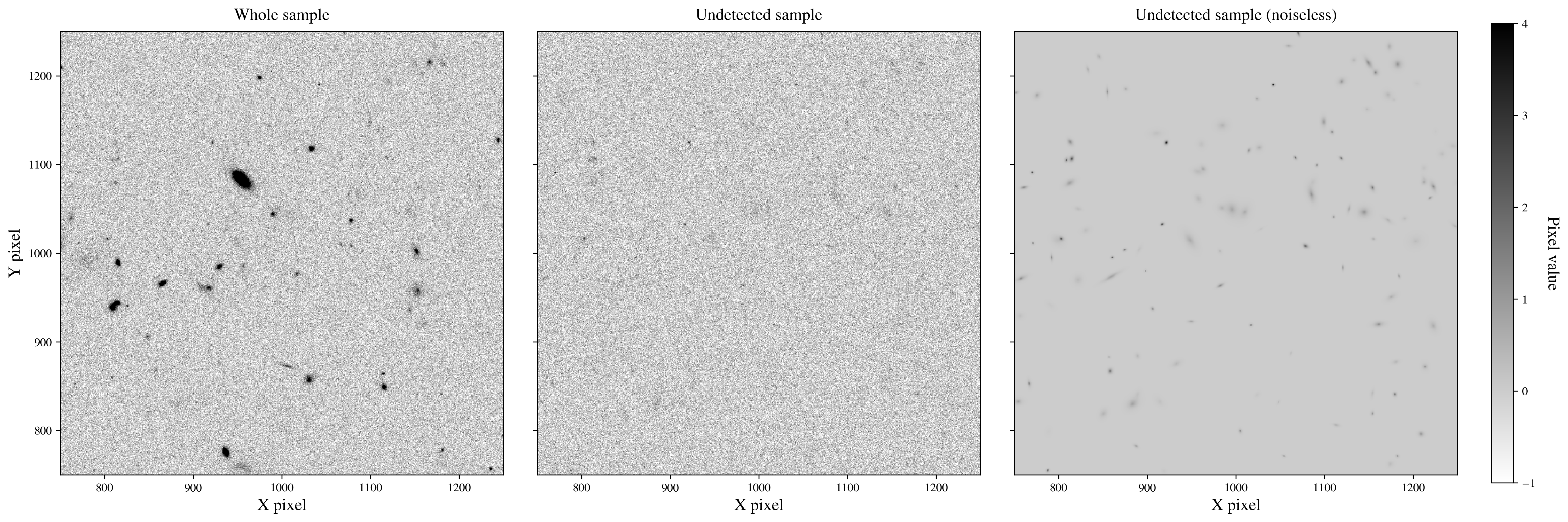}
      \caption{Comparison of simulated \textit{Euclid} VIS images before (left) and after (middle) the removal of detectable galaxies. For clearer visualisation of the clustering of undetected galaxies, the right panel shows a noise-free image of the undetected galaxies alone. The plotted region spans 0.69\,arcmin$^2$, containing 154 galaxies in total, of which 119 remain undetected. Within this region, the brightest galaxy has a VIS input magnitude of 21.0\,mag, compared with 25.2\,mag for the brightest undetected galaxy.}
         \label{fig:ImageComp}
  \end{figure*}

Our input synthetic galaxy populations were drawn from the public Flagship galaxy mock catalogue~\citep{Castander2025AA...697A...5E}, hosted on CosmoHub~\citep{Carretero2017ehep.confE.488C,TALLADA2020100391}. This mock catalogue is built upon the Flagship 2 $N$-body simulation, which employs a simulation box of $3600\ h^{-1}{\rm Mpc}$ per side containing $16000^3$ particles, resulting in a particle mass of $m_{\text{p}} = 10^9\ h^{-1}{\rm M}_{\odot}$. It adopts the \textit{Euclid} reference cosmology: a matter density of $\Omega_{\text{m}} = 0.319$, a baryon density of $\Omega_{\text{b}} = 0.049$, a radiation density of $\Omega_{\text{r}} = 5.509\times 10^{-5}$, and a dark energy density of $\Omega_{\Lambda} = 0.681 - \Omega_{\text{r}} - \Omega_{\nu}$, where the contribution from massive neutrinos is $\Omega_{\nu} = 1.40343\times 10^{-3}$, derived from the minimum possible neutrino mass of 0.0587 eV under the normal hierarchy. The dimensionless Hubble parameter is $h = 0.67$, the scalar spectral index of initial fluctuations is $n_{\text{s}} = 0.96$, and the primordial power spectrum amplitude is $A_{\text{s}} = 2.1 \times 10^{-9}$ at the pivot scale $k_{\text{pivot}} = 0.05\ \text{Mpc}^{-1}$.

Dark matter haloes were identified directly from the lightcone particle data using the Robust Overdensity Calculation using K-Space Topologically Adaptive Refinement (ROCKSTAR) halo finder \citep{Behroozi2013ApJ...762..109B}, resolving structures down to a mass of $10^{11}\ h^{-1}M_{\odot}$. The resulting catalogue contains 16 billion haloes, covering one octant of the sky up to a redshift of $z = 3$. Mock galaxies were subsequently populated into these haloes using a combination of halo occupation distribution and abundance-matching techniques. Free parameters governing this matching process were calibrated against observed scaling relations of various galaxy properties, following the methodology of \citet{Carretero2015MNRAS.447..646C}. The final synthetic galaxy population is complete down to a magnitude limit of $H < 26$ and comprises 3.4 billion galaxies in total. For our study, we utilised the ten-band photometry corresponding to the \textit{Euclid} and LSST filters, the three-dimensional spatial positions of the galaxies, and their morphologies as parameterised by a double S\'ersic profile.

\subsection{Image simulation}

\begin{table*}
    \centering
    \caption{Statistical properties of the simulated \textit{Euclid} and LSST image tiles.}
    \label{tab:sim_image_stats}
    \small
    \renewcommand{\arraystretch}{1.0}
    \begin{tabular}{llccccc}
        \hline
        \hline
        Survey & Band & Pixel scale & Depth & Moffat FWHM & Moffat index & Airy $\lambda_{\text{ref}}$ \\
        & & (arcsec) & (mag) & (arcsec) & & (nm) \\
        \hline
        \multirow{4}{*}{\textit{Euclid}}
        & VIS & 0.1 & $25.70\pm0.10$ & -- & -- & 800 \\
        & $Y$ & 0.3 & $24.06\pm0.08$ & -- & -- & 1080.9 \\
        & $J$ & 0.3 & $24.21\pm0.09$ & -- & -- & 1367.3 \\
        & $H$ & 0.3 & $24.18\pm0.08$ & -- & -- & 1771.4 \\
        \hline
        \multirow{6}{*}{LSST}
        & $u$ & 0.2 & $25.82\pm0.09$ & $0.81\pm0.06$ & 2.08 & -- \\
        & $g$ & 0.2 & $27.05\pm0.09$ & $0.77\pm0.10$ & 2.09 & -- \\
        & $r$ & 0.2 & $27.15\pm0.09$ & $0.73\pm0.05$ & 2.22 & -- \\
        & $i$ & 0.2 & $26.41\pm0.09$ & $0.71\pm0.05$ & 2.46 & -- \\
        & $z$ & 0.2 & $25.61\pm0.09$ & $0.69\pm0.07$ & 2.69 & -- \\
        & $y$ & 0.2 & $24.38\pm0.09$ & $0.68\pm0.07$ & 2.81 & -- \\
        \hline
        \hline
    \end{tabular}
    \tablefoot{The reported depth corresponds to a $5\sigma$ limit with a 2-arcsec diameter aperture. For parameters with uncertainties, the $\pm$ values indicate the $1\sigma$ tile-to-tile Gaussian variation resulting from random draws based on empirical \textit{Euclid} Q1 data quality and LSST ten-year observational expectations. The Moffat indices for the LSST PSFs are adopted from the median values of corresponding bands in the Kilo-Degree Survey, representative of typical ground-based conditions~\citep{Li2023AA...670A.100L}. The \textit{Euclid} PSF is modelled as an Airy profile for a telescope with a 1.2~m diameter and a central obscuration of 0.3, where the reference wavelength ($\lambda_{\text{ref}}$) for each band corresponds to the central wavelength of its respective passband.}
\end{table*}

We simulated noisy images representing the expected synergy between the \textit{Euclid} and LSST surveys using the \texttt{MultiBand\_ImSim} pipeline\footnote{\url{https://github.com/KiDS-WL/MultiBand_ImSim}}~\citep{Li2023AA...670A.100L}, which is built upon the public \texttt{GalSim} package\footnote{\url{https://github.com/GalSim-developers/GalSim}}~\citep{Rowe2015AC....10..121R}. These two Stage-IV photometric surveys share a substantial observational footprint, prompting active discussions regarding their combined scientific potential~\citep[e.g.][]{Rhodes2017ApJS..233...21R,Capak2019arXiv190410439C,Guy2022zndo...5836022G}. Our simulations mimic the overlapping sky regions observed by the nominal wide-field surveys of both instruments, covering the \textit{Euclid} visible instrument (VIS) band and Near-Infrared Spectrometer and Photometer (NISP) $Y$, $J$, and $H$ bands, alongside the LSST $u$, $g$, $r$, $i$, $z$, and $y$ bands.

To reflect the anticipated imaging quality accurately, we derived the \textit{Euclid} survey depth information from the \textit{Euclid} Quick Data Release 1 (Q1) products, which share the same imaging quality as the \textit{Euclid} nominal wide-field survey~\citep{Euclid2025arXiv250315302E}. As the nominal LSST survey data are not yet publicly available, we relied on the ten-year characteristic expectations~\citep{Bianco2022ApJS..258....1B}. The PSF is modelled as an Airy profile for the space-based \textit{Euclid} images and a Moffat profile for the ground-based LSST images. For both surveys, we assumed the background noise follows a Gaussian distribution, corresponding to data products that have been accurately corrected for systematic instrumental effects. In practice, astrophysical foreground contamination, such as Galactic cirrus, would also contribute to the background; we leave this more realistic simulation scenario for future work. Since our cosmological analyses focus primarily on fluctuations measured through higher-order statistics, such as two-point correlation functions, these smoother contaminating signals would enter only as higher-order corrections.

To introduce realistic observational variations across the survey footprint, arising from pointing-dependent light contamination (predominantly zodiacal light for \textit{Euclid}) and varying night conditions for LSST, we simulated image tiles of one square degree each. For every tile, the background noise level and the PSF full width at half maximum (FWHM) were independently drawn from Gaussian distributions. The parameters of these distributions were informed either by the empirical variations measured in the \textit{Euclid} Q1 data or, in the case of LSST, by representative expectations that account for the impact of potentially missing exposures~\citep{Bianco2022ApJS..258....1B}. The statistical properties of the resulting simulated images are summarised in Table~\ref{tab:sim_image_stats}.

For the training sample, we simulated 100 tiles, corresponding to a total area of 100 square degrees drawn from the Flagship simulation. Model performance is evaluated on an independent 50-square-degree region, for which statistically independent noise and PSF realisations were generated. In addition to evaluation samples drawn from the same nominal distributions as the training data, we generated two out-of-distribution test sets in which the background noise level was systematically shifted by $\pm3\sigma$, respectively. Importantly, each of these sets contains a distinct underlying undetected galaxy sample: as detailed in Section~\ref{Sec:unresolved}, the undetected sample construction procedure was performed independently for every noise configuration, ensuring that changes in noise level propagate consistently into the selection of the undetected population. These controlled perturbations allow us to quantify the sensitivity of the model to mismatches in the background noise level.

\subsection{Construction of undetected galaxy samples}
\label{Sec:unresolved}

Techniques for extracting the unresolved extragalactic background light continue to evolve, ranging from simple masking approaches \cite[e.g.][]{Lim2023MNRAS.525.1443L} to more sophisticated light-subtraction methods~\citep[e.g.][]{Windhorst2022AJ....164..141W}. To circumvent the complexities of these techniques, we opted to simulate the unresolved background directly by identifying and removing all detectable galaxies from our input catalogue. This strategy reduces the required simulation volume and, under the assumption of perfect masking or subtraction, effectively mimics either the unmasked regions in a masking approach or the residual images produced by light-subtraction pipelines.

To identify detectable galaxies in our input catalogue, we performed two iterations of VIS image simulation and source detection with \textsc{SExtractor}~\citep{Bertin1996AAS..117..393B}, using the configuration of \citet{Hoekstra2021AA...646A.124H}. After each iteration, we removed sources from the input catalogue that matched entries in the detected catalogue. Repeating this procedure mitigates cross-matching failures caused by blended galaxies being detected as a single source. Because blends involving more than two galaxies can still persist after the second iteration, we additionally imposed VIS$>$24 to remove any remaining bright galaxies.

\begin{figure}
  \centering
  \includegraphics[width=\hsize]{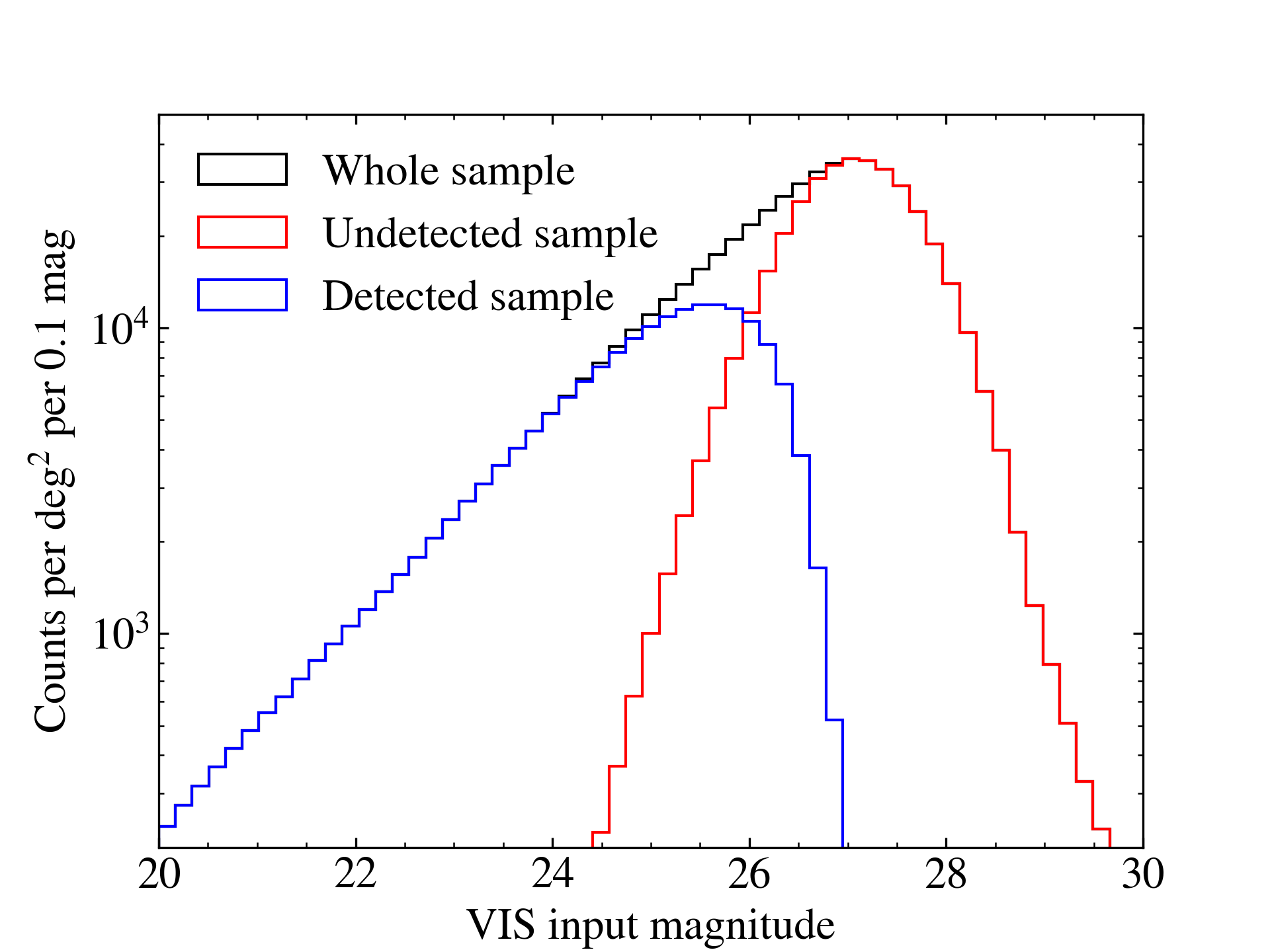}
      \caption{Comparison of the \textit{Euclid} VIS magnitude distributions of the whole sample (black), the identified detectable sample (blue), and the undetected sample (red). The overall decline in source counts beyond a magnitude of ${\sim}27$ reflects the intrinsic mass resolution limits of the Flagship simulation.}
         \label{fig:InputMag}
  \end{figure}

Figure~\ref{fig:ImageComp} compares the simulated \textit{Euclid} VIS images before and after the removal of detectable galaxies. The corresponding input VIS magnitude distributions are compared in Figure~\ref{fig:InputMag}. As shown, the majority of galaxies are detected by \textit{Euclid} down to a VIS magnitude of ${\sim}25$. Beyond this, detection completeness drops steadily, with almost no sources detected fainter than a magnitude of ${\sim}26.5$. This behaviour is consistent with the expected performance of \textit{Euclid}.

The decline in galaxy counts beyond VIS magnitude ${\sim}27$ reflects the finite depth of the input mock catalogue built from the Flagship simulations and represents a fundamental limitation of our current study. Real unresolved-background maps contain additional fainter galaxies, which can modify both the multi-band pixel fluxes and the underlying redshift distributions. Their impact on our main target, the flux-weighted redshift distributions, depends on their integrated flux, colours, redshift distributions, and spatial clustering. If the additional population contributes appreciable flux with similar colours and redshift distributions across pixels, it could make the pixels less distinguishable and thus weaken tomographic separation. The extent of this dilution, however, depends on how the fainter population traces large-scale structure, since its spatial clustering may preserve redshift-dependent colour variations between pixels and thereby retain information useful for tomography. Therefore, quantifying the net impact of the missing faint population requires extending our simulations to include these galaxies with realistic photometric properties and spatial distributions.

\subsection{Projection of images onto HEALPix maps}

  \begin{figure*}
  \centering
  \includegraphics[width=0.85\textwidth]{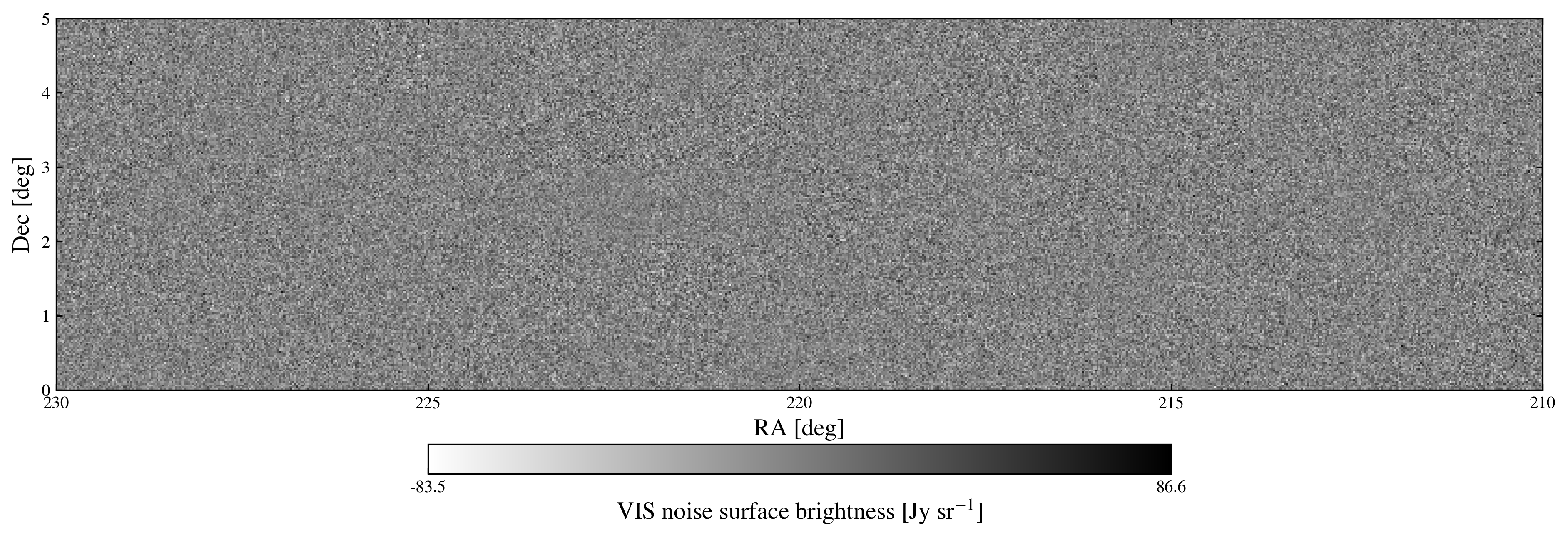}\par
  \includegraphics[width=0.85\textwidth]{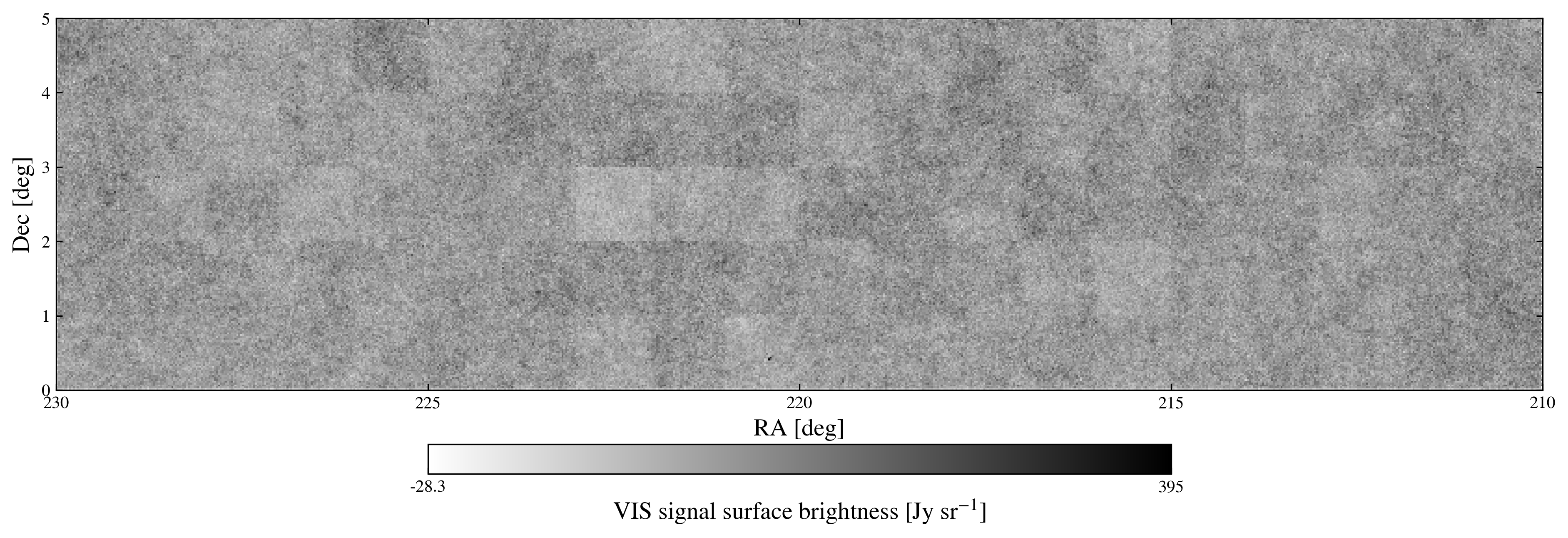}\par
  \includegraphics[width=0.85\textwidth]{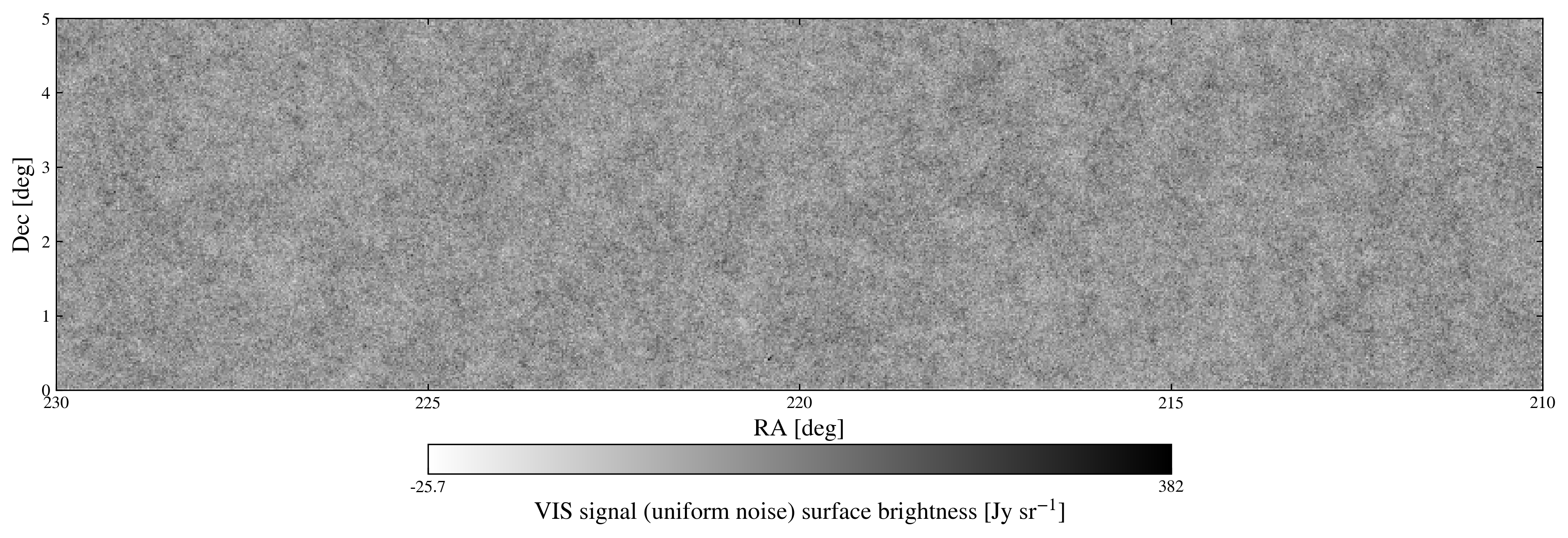}\par
      \caption{Comparison of HEALPix maps constructed from the simulated \textit{Euclid} VIS images, without (top) and with (middle) undetected galaxies. The prominent tile-level variations arise from changing background noise levels across the footprint, whilst the finer, smaller-scale fluctuations reveal the underlying large-scale structure traced by the undetected galaxies. The bottom panel shows the same map with a uniform noise level across all tiles, in which the large-scale structure is more clearly visible. This uniform-noise case is presented here solely for visual comparison and is not used in model training.}
         \label{fig:HealpixMap}
  \end{figure*}

The simulated images containing only the undetected galaxies were subsequently transformed into Hierarchical Equal Area iso-Latitude Pixelisation (HEALPix) maps using the \texttt{healpy} package\footnote{\url{https://github.com/healpy/healpy}} \citep{Gorski2005ApJ...622..759G, Zonca2019}. We adopted an $N_{\text{side}}$ of 4096, corresponding to an angular pixel scale of approximately 0.86 arcminutes. Under our nominal noise configurations, this resolution yields an average of ${\sim}130$ undetected galaxies per HEALPix pixel. The value of each pixel was calculated by summing the flux of all original simulated image pixels enclosed within its boundaries and then dividing by the solid angle of the HEALPix pixel to obtain a surface brightness in units of janskys per steradian (Jy\,sr$^{-1}$). By integrating the signal over a larger solid angle, we effectively suppress uncorrelated instrumental noise, thereby revealing the underlying large-scale structure traced by the undetected background galaxies.

Figure~\ref{fig:HealpixMap} provides a visual comparison of the HEALPix maps constructed from the \textit{Euclid} VIS images, with and without undetected galaxies. Clustering features are clearly visible, superimposed on the tile-level variations caused by the background noise fluctuating between tiles. For visual clarity, we also include a map with a uniform noise level across all tiles, in which the large-scale structure traced by the undetected galaxies is more clearly visible. We note that this uniform-noise map is presented solely for visual comparison and is not used in model training. As a first-order check of the signal significance, we computed the pixel-to-pixel standard deviation of the surface brightness across the maps. The pure noise map yields a root-mean-square (RMS) fluctuation of 17.98 Jy\,sr$^{-1}$ (surface brightness values in Jy\,sr$^{-1}$ convert to AB\,mag\,arcsec$^{-2}$ via $\mu_{\rm AB}= -2.5\log_{10}({S_\nu}/[\mathrm{Jy}\,\mathrm{sr}^{-1}])+35.47$, where the constant follows from the AB zero point of 3631\,Jy and $1\,\rm sr\approx4.255\times10^{10}\,arcsec^2$), whereas the map containing the undetected galaxies exhibits an increased RMS of 34.11 Jy\,sr$^{-1}$.

The per-pixel RMS, $\sigma_{\rm pix}$, can be scaled to any aperture $A_{\rm ap}$ using the HEALPix pixel area of $\Omega_{\rm pix}\approx2656\,{\rm arcsec}^2$ at $N_{\rm side}=4096$, with the effective noise given by $\sigma_{\rm ap}=\sigma_{\rm pix}\sqrt{\Omega_{\rm pix}/A_{\rm ap}}$. Applying this scaling, the noise map RMS corresponds to a $3\sigma$ limiting surface brightness of 29.4\,mag\,arcsec$^{-2}$ within a $10\times10\ {\rm arcsec}^2$ area. This is remarkably close to the expected \textit{Euclid} Wide Survey limit of $29.5^{+0.08}_{-0.27}$\,mag\,arcsec$^{-2}$ for detecting extended low-surface-brightness structures, as predicted by the more comprehensive analysis in \citet{Euclid2022AA...662A.112E,Euclid2022AA...657A..92E}. This excellent agreement demonstrates that, despite the simplified noise model employed in our simulations, our maps faithfully represent the expected real-world performance of the \textit{Euclid} VIS instrument.

\begin{figure}
  \centering
  \includegraphics[width=\hsize]{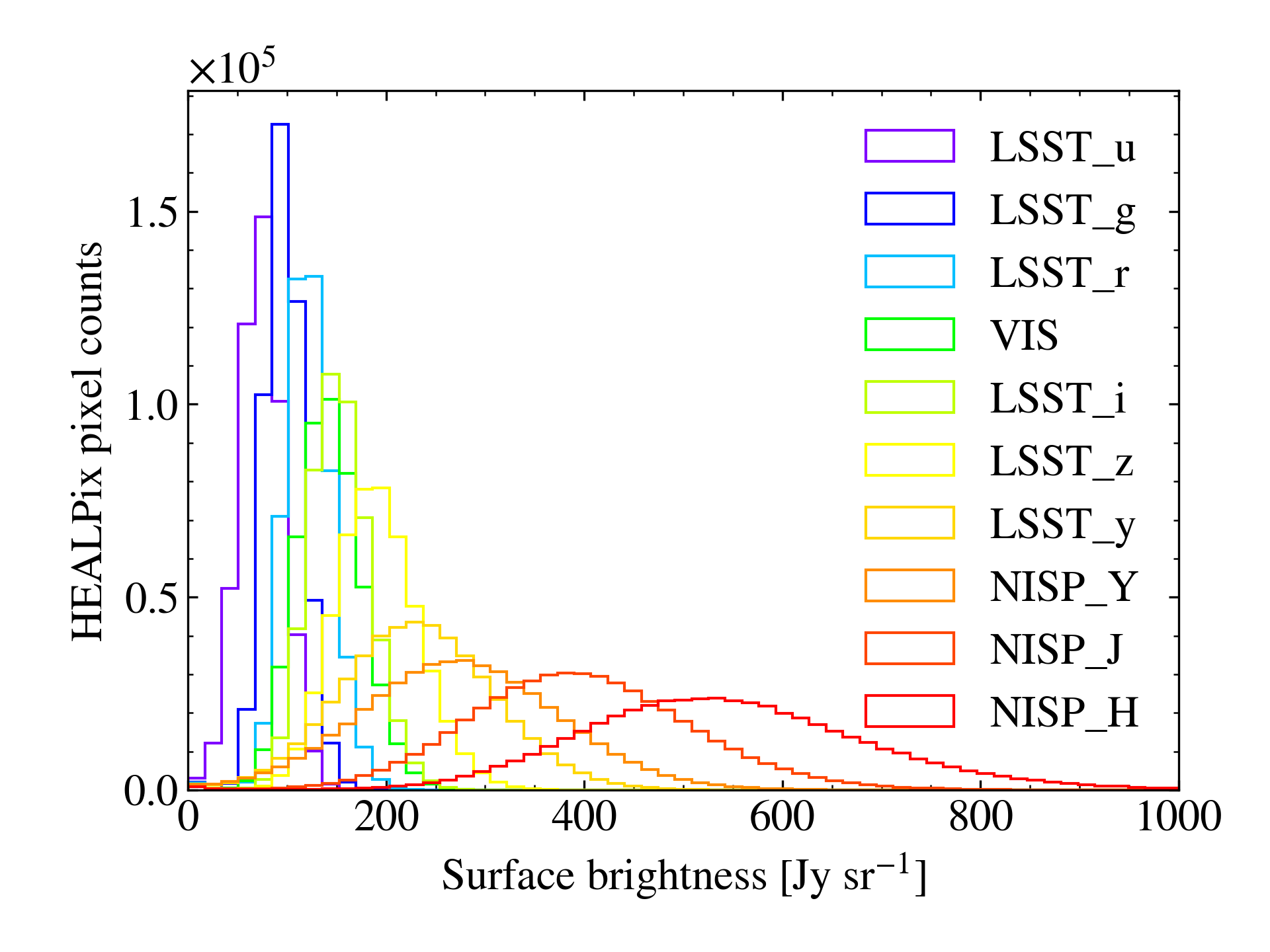}
      \caption{Distributions of pixel values across all ten bands of the HEALPix maps in the training sample. A general trend towards higher overall surface brightness and broader distributions is observed in the redder bands.}
         \label{fig:HealpixHist}
  \end{figure}

Figure~\ref{fig:HealpixHist} shows the pixel value distributions across all ten bands of the HEALPix maps in our training sample. Variations in the flux distributions are evident between bands, with a general trend towards higher overall surface brightness and broader distributions in the redder bands. These distinct statistical characteristics encode the essential broad-band colour information that enables inference of the redshift distribution and other intrinsic properties of the undetected galaxy population.

\section{Methods}
\label{Sec:method}

Estimating galaxy redshifts from broad-band colours, known as photometric redshift estimation, is a well-established technique in modern wide-field imaging surveys (see \citealt{Newman2022ARAA..60..363N} for a recent review). Methods broadly fall into two categories: template-fitting methods, which use models of galaxy spectral energy distributions, and empirical methods, which learn the relationship between observed colours and redshifts. Our framework extends the empirical approach to undetected galaxies using conditional normalising flows. These models learn probability distributions by transforming a simple base distribution, such as a standard normal distribution, into a complex target distribution through a sequence of invertible, parameterised transformations. Incorporating external conditioning variables allows the model to learn how the target distribution depends on the observed inputs. The framework also naturally accommodates joint, multi-dimensional target distributions, enabling us to capture dependences among galaxy properties required for physical signal modelling and to include observable diagnostics for assessing model reliability and uncertainty. Appendix~\ref{Sec:flow} provides more details on this technique, with a schematic diagram of the overall architecture shown in Figure~\ref{fig:flow_schematic}.

\subsection{Variable specification and data preprocessing}

In this exploratory study, we used a normalising flow to jointly predict the distributions of three quantities characterising the undetected galaxy population --- its redshift, VIS magnitude, and the RMS of the VIS imaging noise --- conditioned on features derived from the multi-band pixel values. The VIS magnitude is included as one of the three target quantities to facilitate the estimation of VIS flux-weighted redshift distributions, which are required for modelling the VIS band maps of unresolved background light. Depending on the specific band under analysis, magnitudes from other photometric bands can readily be substituted into this framework. The RMS of the VIS imaging noise is included to aid the assessment of model uncertainties, particularly in scenarios where the noise properties of the training sample differ from those of the target data. This approach can naturally be extended to incorporate other observational systematics, addressing a critical yet frequently overlooked aspect of uncertainty quantification in machine learning and simulation-based inference.

The conditioning variables were derived from the pixel values of the ten-band HEALPix maps. Because photometric redshift correlates more strongly with galaxy colours than with individual band fluxes, we constructed our conditioning inputs using the flux ratios between adjacent bands, supplemented by the VIS-band flux, yielding a ten-dimensional feature space. Prior to training, both the three-dimensional target and the ten-dimensional conditioning variables were standardised to have a zero mean and unit variance, ensuring stable and efficient convergence.

\subsection{Model specification and training procedure}
\label{Sec:training}

For the model architecture, we employed coupling layers with monotonic rational–quadratic splines as the transformer \citep[introduced in Appendix~\ref{Sec:flow}; see also][]{Dinh2014arXiv1410.8516D,Durkan2019arXiv190604032D}, as implemented in the \texttt{FlowJAX} package (v17.2.1)\footnote{\url{https://github.com/danielward27/flowjax}}. The base distribution was defined as a three-dimensional standard normal distribution, matching the dimensionality of the target space. To ensure that all dependences across the target dimensions are comprehensively captured, we stacked eight coupling layers interleaved with random permutations. The boundary for the spline transformer was set to $[-5, 5]$. Given our standardised data preprocessing, this range corresponds approximately to the $5\sigma$ extent of the distribution, rendering the fraction of outliers beyond these bounds negligible. The number of spline knots was set to 16 to ensure smooth transformations within this interval. For the conditioning network, we adopted a multi-layer perceptron with four hidden layers, each with a width of 256 units, and rectified linear unit activation functions.

For the training procedure, we optimised the network by minimising the negative log-likelihood of the target distribution conditioned on the adjacent-band flux ratios plus the VIS-band flux. The model was trained using the Adam optimiser with a learning rate of $10^{-5}$. To ensure stable convergence and prevent exploding gradients, we applied gradient clipping, restricting the global gradient norm to a maximum value of one. The training data were processed in mini-batches of 8192 samples. To monitor model generalisation, we randomly reserved 20\% of the training dataset as an internal validation set. We note that this internal validation set is distinct from the final evaluation datasets, as described in Section~\ref{Sec:Sim}. Training concluded after 660 epochs, when the early-stopping criterion was met with a patience of 40 epochs, taking approximately nine hours on a single graphics processing unit. The final loss values for the training set (2.08) and the internal validation set (2.01) are closely matched, indicating that the model is well fitted with no evidence of significant over-fitting.

\begin{table*}
    \centering
    \caption{Summary statistics for model performance across various test scenarios.}
    \label{tab:summary}
    \small
    \renewcommand{\arraystretch}{1.0}
    \begin{tabular}{lcccc|cccc}
        \hline
        \hline
        \multirow{2}{*}{Test sample} & \multicolumn{4}{c}{Redshift density} & \multicolumn{4}{c}{VIS flux density} \\
        \cline{2-5} \cline{6-9}
        & $\Delta\mu$ & $\Delta\sigma$ & EMD & KS & $\Delta\mu$ & $\Delta\sigma$ & EMD & KS \\
        & $(10^{-2})$ & $(10^{-2})$ & $(10^{-2})$ & $(10^{-2})$ & $(10^{-9})$ & $(10^{-9})$ & $(10^{-9})$ & $(10^{-2})$\\
        \hline
        Pure noise                    & $-43.8$ & $+8.66$ & $+43.5$ & $+28.0$ & $+58.6$ & $+74.9$ & $+46.2$ & $+91.1$ \\
        Fiducial                      & $+0.15$ & $+0.60$ & $+0.61$ & $+0.65$ & $-0.22$ & $-0.97$ & $+0.79$ & $+0.79$ \\
        20\% missing (zero)           & $-8.70$ & $+6.51$ & $+8.26$ & $+6.98$ & $+2.96$ & $+24.3$ & $+10.2$ & $+13.4$ \\
        Missing LSST $y$ (zero)       & $-1.06$ & $-1.29$ & $+2.11$ & $+1.92$ & $-5.58$ & $-2.69$ & $+6.76$ & $+5.58$ \\
        Missing all LSST (zero)       & $+34.1$ & $+3.64$ & $+34.4$ & $+22.7$ & $-26.7$ & $-21.2$ & $+26.2$ & $+31.8$ \\
        Shift RMS $-3\sigma$          & $-2.70$ & $+1.41$ & $+2.31$ & $+1.80$ & $+6.33$ & $+8.73$ & $+4.18$ & $+4.96$ \\
        Shift RMS $+3\sigma$          & $+1.58$ & $+0.21$ & $+1.98$ & $+1.50$ & $-6.71$ & $-11.0$ & $+4.83$ & $+3.87$ \\
        \hline
        & \multicolumn{4}{c}{VIS noise RMS density} & \multicolumn{4}{c}{VIS flux-weighted redshift density} \\
        \cline{2-5} \cline{6-9}
        & $\Delta\mu$ & $\Delta\sigma$ & EMD & KS & $\Delta\mu$ & $\Delta\sigma$ & EMD & KS \\
        & $(10^{-11})$ & $(10^{-11})$ & $(10^{-11})$ & $(10^{-2})$ & $(10^{-2})$ & $(10^{-2})$ & $(10^{-2})$ & $(10^{-2})$\\
        \hline
        Pure noise                    & $-1.95$ & $+12.6$ & $+12.7$ & $+25.2$ & $-56.6$ & $+7.94$ & $+56.3$ & $+35.5$ \\
        Fiducial                      & $-0.48$ & $+2.65$ & $+2.39$ & $+5.19$ & $+0.80$ & $-0.31$ & $+0.86$ & $+0.74$ \\
        20\% missing (zero)           & $-3.61$ & $+9.59$ & $+9.32$ & $+24.1$ & $-11.4$ & $+4.81$ & $+11.1$ & $+7.97$ \\
        Missing LSST $y$ (zero)       & $-12.0$ & $+1.73$ & $+12.4$ & $+37.9$ & $+0.73$ & $-1.71$ & $+2.36$ & $+1.76$ \\
        Missing all LSST (zero)       & $+2.00$ & $+3.17$ & $+7.09$ & $+33.0$ & $+43.7$ & $+3.79$ & $+43.7$ & $+25.9$ \\
        Shift RMS $-3\sigma$          & $+47.0$ & $+5.42$ & $+45.9$ & $+86.9$ & $-3.64$ & $+0.10$ & $+3.64$ & $+2.54$ \\
        Shift RMS $+3\sigma$          & $-53.0$ & $+1.21$ & $+52.2$ & $+89.8$ & $+3.67$ & $-0.36$ & $+3.67$ & $+2.41$ \\
        \hline
    \end{tabular}
    \tablefoot{The four evaluation metrics presented are the differences in the distribution mean ($\Delta\mu$) and standard deviation ($\Delta\sigma$), the Earth Mover's Distance (EMD), and the Kolmogorov--Smirnov (KS) statistic. Values are scaled by the factors indicated in the column headers. The fiducial case represents the benchmark performance under idealised conditions, whilst the pure-noise case provides an empirical baseline for assessing model performance. Detailed descriptions of all test samples are provided in Section~\ref{Sec:res}. For the missing data scenarios, the quoted values reflect raw model performance prior to any data imputation; as discussed in the text, applying our imputation strategies restores performance to near the fiducial level.}
\end{table*}

\section{Results}
\label{Sec:res}

  \begin{figure*}
  \sidecaption
  \includegraphics[width=12cm]{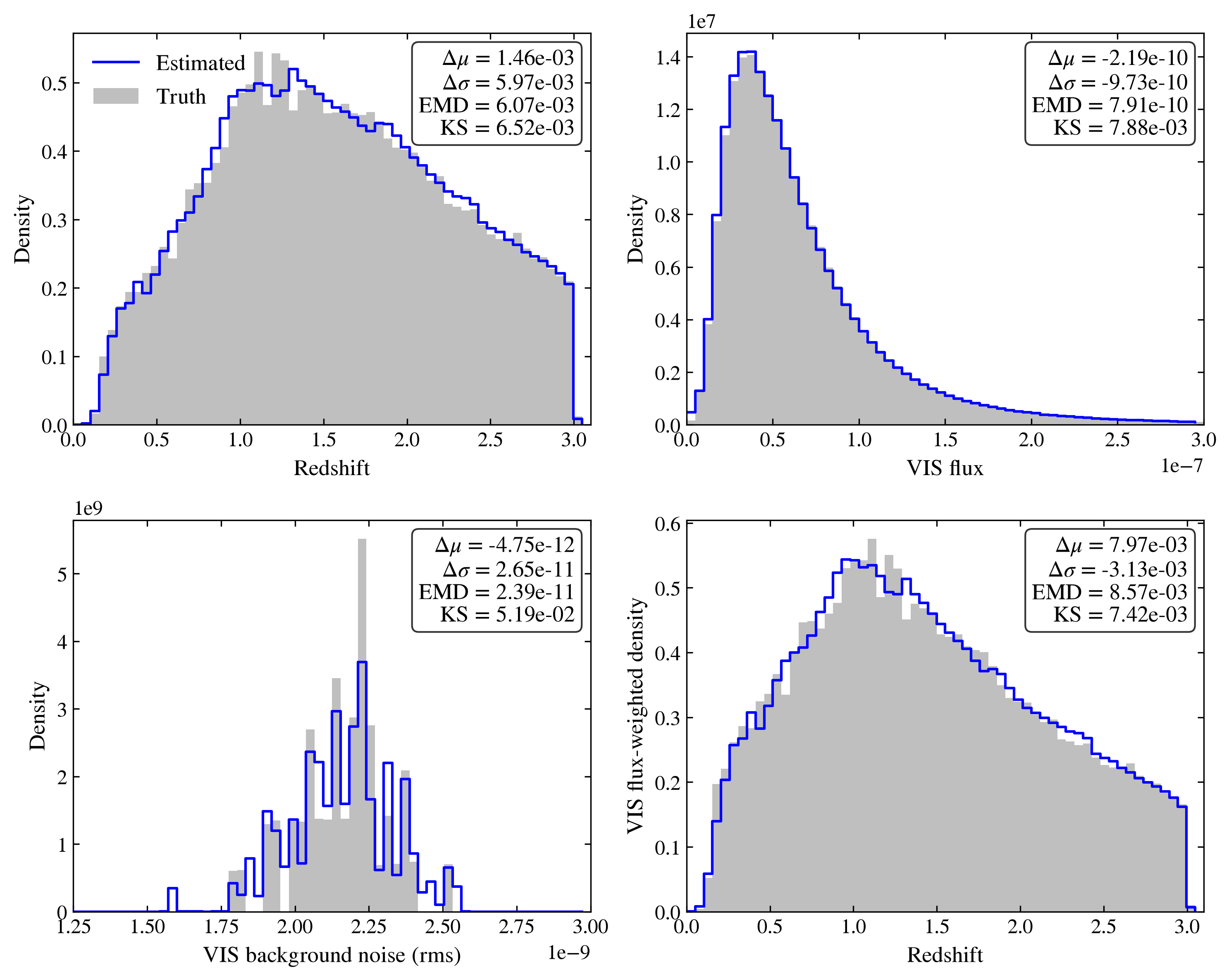}
      \caption{Comparison of the model-predicted target distributions (blue lines) with the underlying truth (shaded histograms) for the fiducial evaluation sample. The four panels, in clockwise order, correspond to the redshift distribution, VIS flux distribution, VIS flux-weighted redshift distribution, and VIS background noise RMS distribution. The quoted summary statistics are also presented in Table~\ref{tab:summary}.}
         \label{fig:fiducial}
  \end{figure*}

Figure~\ref{fig:fiducial} compares the model-predicted target distributions with the underlying truth, evaluated on the independent fiducial test sample described in Section~\ref{Sec:Sim}. We emphasise that this evaluation set is drawn from a different sky region and therefore contains a distinct galaxy population, as well as independently generated noise properties, relative to the training data. The predicted distributions show excellent agreement with the ground truth, demonstrating the model’s ability to recover the desired galaxy properties from multi-band map observations.

To quantify model performance more precisely, we computed several summary statistics comparing the model-predicted target distributions with the true ones: offsets in the distribution mean ($\Delta\mu$) and standard deviation ($\Delta\sigma$), the earth mover's distance (EMD, or Wasserstein-1 metric, \citealt{Ramdas2015arXiv150902237R}), and the Kolmogorov--Smirnov (KS) statistic \citep{kolmogorov1933sulla, smirnov1939estimation}. The first two are the quantities commonly adopted as calibration requirements in current galaxy-based analyses. For example, weak lensing studies require sub-per cent level accuracy in $\Delta\mu$ in order to meet the cosmological constraining power targeted by upcoming galaxy surveys \citep{Ma2006ApJ...636...21M,Euclid2026arXiv260400805E}. The latter two are distribution-level metrics, included to quantify how well our algorithm recovers the finer details of the distributions, which can become important depending on the statistical power of the survey and the cosmological probe considered \citep{Schmidt2020MNRAS.499.1587S,Stolzner2021AA...650A.148S,Cawthon2022MNRAS.513.5517C,Euclid2026arXiv260400805E}.

The EMD, defined as $W_1(p, q)=\int\,|F_{p}(x)-F_{q}(x)|\, {\rm d}x$, where $F_p$ and $F_q$ are the cumulative distribution functions of the true distribution $p$ and the predicted distribution $q$, measures the minimum `cost' of transforming one distribution into the other, and is particularly responsive to systematic shifts in the bulk of the distribution. Because it carries the units of the variable itself, it can be read directly as the effective offset between the predicted and true distributions. The KS statistic, defined as $D_{\rm KS}(p, q)=\max_{x}|F_{p}(x)-F_{q}(x)|$, instead records the single largest discrepancy between the two cumulative distributions, and is therefore sensitive to localised disagreement that the integrated EMD can average away. It is dimensionless and bounded between 0 and 1, and can be read directly as the maximum difference in cumulative probability between the two distributions.

To provide an empirical reference for all four statistics, we additionally applied the model to a pure-noise sample containing only background noise and no galaxies. This null test provides a no-information reference against which the values measured on the real test samples can be judged. We emphasise, nonetheless, that these statistics characterise only how closely the predicted distributions match the true ones; translating them fully into requirements for cosmological inference would require dedicated downstream forecasts of the kind performed for galaxy surveys \citep{Ma2006ApJ...636...21M,Euclid2026arXiv260400805E}. We leave such forecasts to future work, and quote these values as an indication of the general performance of our model in recovering the target distributions.

The resulting metric values are reported in the corresponding figures and summarised in Table~\ref{tab:summary}. The model yields substantially smaller discrepancies than the null test across all metrics for the fiducial sample. In particular, the primary result of VIS flux-weighted redshift density estimation reaches sub-per cent accuracy in both the mean and standard deviation. This level of agreement demonstrates the model’s ability to capture the complex dependences between the conditioning and target distributions, thereby validating the feasibility of our modelling approach.

Nevertheless, we emphasise that this fiducial performance corresponds to an idealised scenario in which the training sample is statistically well matched to the target sample in both imaging conditions and the underlying galaxy population. Such ideal conditions are rarely met in real-world surveys. To address this, the following sections assess model robustness against common observational imperfections, introduce prediction-based diagnostics to identify and potentially correct these mismatches, and explore the feasibility of tomographic analyses using the model predictions.

\subsection{Impact of incomplete photometric coverage}
\label{Sec:incomplete}

\begin{figure}
  \centering
  \includegraphics[width=\hsize]{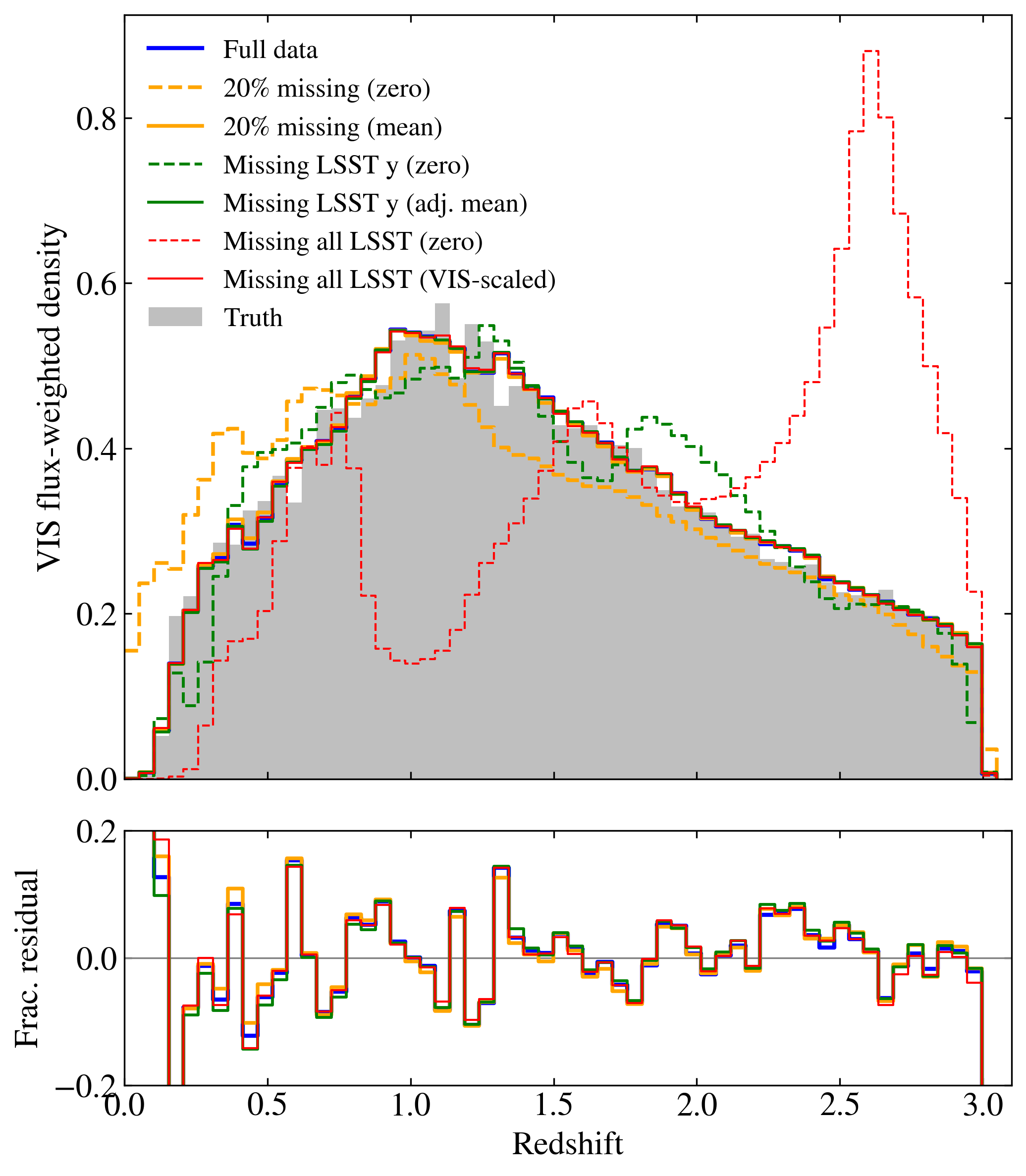}
      \caption{Comparison of VIS flux-weighted redshift distributions between model predictions (lines) and the underlying truth (shaded histogram) for different levels of missing data. Top: Distributions. Bottom: Fractional residuals, defined as $(n_{\mathrm{pred}} - n_{\mathrm{true}}) / n_{\mathrm{true}}$. Different colours correspond to different levels of missing data, with dashed lines indicating raw results without imputation and solid lines indicating results after applying the imputation strategies detailed in Section~\ref{Sec:incomplete}.}
         \label{fig:incomplete}
  \end{figure}

One of the most common observational systematics arises from missing data in one or more imaging bands. To simulate these effects, we masked portions of the conditioning inputs used for model prediction. Specifically, we considered three representative scenarios: (1) randomly masking 20\% of pixels in each band, (2) removing the entire LSST $y$-band imaging, and (3) an extreme case in which all LSST bands are missing. The corresponding results are shown in Figure~\ref{fig:incomplete}, with quantitative summary statistics detailed in Table~\ref{tab:summary}. Unsurprisingly, model performance degrades significantly when missing data are set to zero, with the most severe impact occurring when all LSST bands are removed, followed by the case of 20\% missing pixels per band.

However, we find that the impact of such observational effects can be largely mitigated through certain data imputation strategies. Rather than assigning zero values to missing data, we can replace them with representative estimates derived from the available observations. For instance, in the 20\% random masking scenario, missing pixels can be imputed using the mean value of the remaining unmasked data within each band. When the LSST $y$ band is missing, the absent data can be approximated by using the average of its adjacent photometric bands. Finally, when all LSST bands are missing, we can extrapolate from the available VIS-band fluxes by scaling them according to the ratio of global mean fluxes between the VIS and the respective target bands.

With these imputation strategies, model performance recovers to a level comparable to the fiducial case, as illustrated by the solid lines in Figure~\ref{fig:incomplete}. This demonstrates that the model is highly robust to incomplete data coverage. This resilience is likely attributable to the fact that conditional normalising flows capture global statistical correlations between the conditioning inputs and the target distributions, rather than depending on fine-grained features. Whilst this generalisation behaviour is advantageous for handling missing observations, it can become a limitation when systematic mismatches exist between the training and target samples, as we evaluate in the following section.

\subsection{Impact of imaging noise mismatches}
\label{Sec:noiseMis}

Perhaps the most demanding challenge for supervised machine learning techniques is the potential mismatch between the training and target samples. Fundamentally, this can only be addressed by increasing the realism of the training sample. However, before any such correction can be applied, one must first be able to diagnose whether a mismatch exists. This is challenging when all properties predicted by the model — such as the redshift and magnitude distributions — are intrinsically unknown from the observations. To address this, we include the VIS background noise RMS as an additional target variable alongside the intrinsic galaxy properties of interest. The rationale is that the background noise RMS is an observable quantity that can be measured from real data, providing a direct point of comparison between model predictions and observations.

\begin{figure*}
  \sidecaption
  \includegraphics[width=12cm]{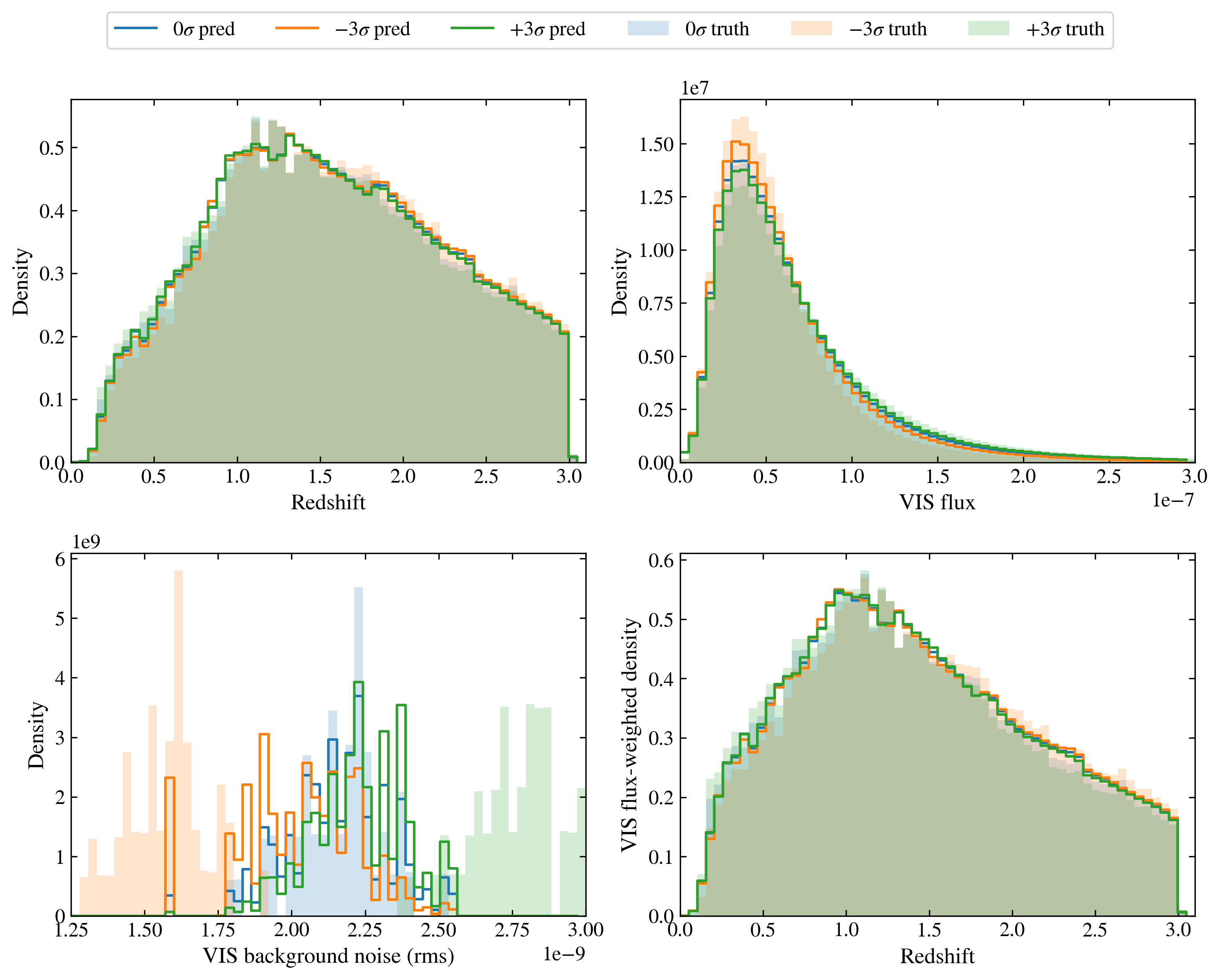}
      \caption{Comparison of the model-predicted target distributions (lines) with the underlying truth (shaded histograms) for the out-of-distribution test samples. The four panels, in clockwise order, correspond to the redshift distribution, VIS flux distribution, VIS flux-weighted redshift distribution, and VIS background noise RMS distribution. Different colours correspond to different levels of shift in the imaging noise background, as detailed in Section~\ref{Sec:Sim}.}
         \label{fig:shift}
  \end{figure*}

To test both the impact of noise mismatches and the effectiveness of this diagnostic, we prepared two out-of-distribution test sets as detailed in Section~\ref{Sec:Sim}. These sets feature systematically shifted background noise levels in the simulated images, which in turn alter the underlying sample composition: more galaxies remain undetected in the higher-noise case, whilst fewer do so in the lower-noise case, as illustrated by the shaded histograms in Figure~\ref{fig:shift}.

We ran model predictions using conditioning variables from these two out-of-distribution test sets, namely the features derived from the ten-band HEALPix pixel values. The results are shown as solid lines in Figure~\ref{fig:shift}, with summary statistics presented in Table~\ref{tab:summary}. Whilst the model does adjust its predictions in response to the changed pixel values, these adjustments are insufficient to fully capture the true shifts in the underlying distributions. This is partly due to the statistical rigidity of normalising flow models discussed earlier, but also because changes in noise level alter the selection of the underlying undetected population — a process that the trained model has no capacity to predict.

Crucially, as indicated in Figure~\ref{fig:shift}, mismatches in the VIS background noise RMS are notably more prominent than deviations in the intrinsic galaxy properties of interest. This confirms the value of including an observable diagnostic quantity among the target variables: the discrepancy in the predicted noise RMS can serve as an effective indicator of model uncertainty, flagging cases where the training sample may not be representative of the target data and guiding the refinement of the training sample for retraining if improved accuracy is required.

\subsection{Feasibility of tomographic analyses}

Tomographic analyses, which split samples into bins based on redshift, are commonly employed in cosmological inference as they provide additional information that allows one to explore the evolution of large-scale structure over cosmic time~\citep{Hu1999ApJ...522L..21H,Huterer2002PhRvD..65f3001H}. In traditional catalogue-based analyses, individual detected galaxies are assigned to tomographic bins using their photometric redshift estimates. In our map-based framework, although individual galaxies are undetected, an analogous procedure can be performed at the pixel level by leveraging the model's per-pixel redshift estimates. This is motivated by the expectation that galaxies within the same pixel occupy a more constrained redshift range than the full sample, due to the spatial clustering of galaxies.

Specifically, the model draws 30 redshift samples per pixel, and the final global redshift distribution is a combination of these individual per-pixel redshift distributions. This provides a natural means of treating each pixel as a fundamental unit — analogously to individual galaxies in a catalogue-based analysis — and assigning it to a tomographic bin based on its mean predicted redshift. To test this, we assigned pixels to four tomographic bins with redshift edges at $z = 0, 1.2, 1.4, 1.6, 3.0$, using the mean of the 30 redshift estimates per pixel. We then aggregated the per-pixel redshift distributions within each bin separately to assess whether the model can recover the true underlying redshift distributions for these tomographically defined subsamples.

\begin{figure*}
  \sidecaption
  \includegraphics[width=12cm]{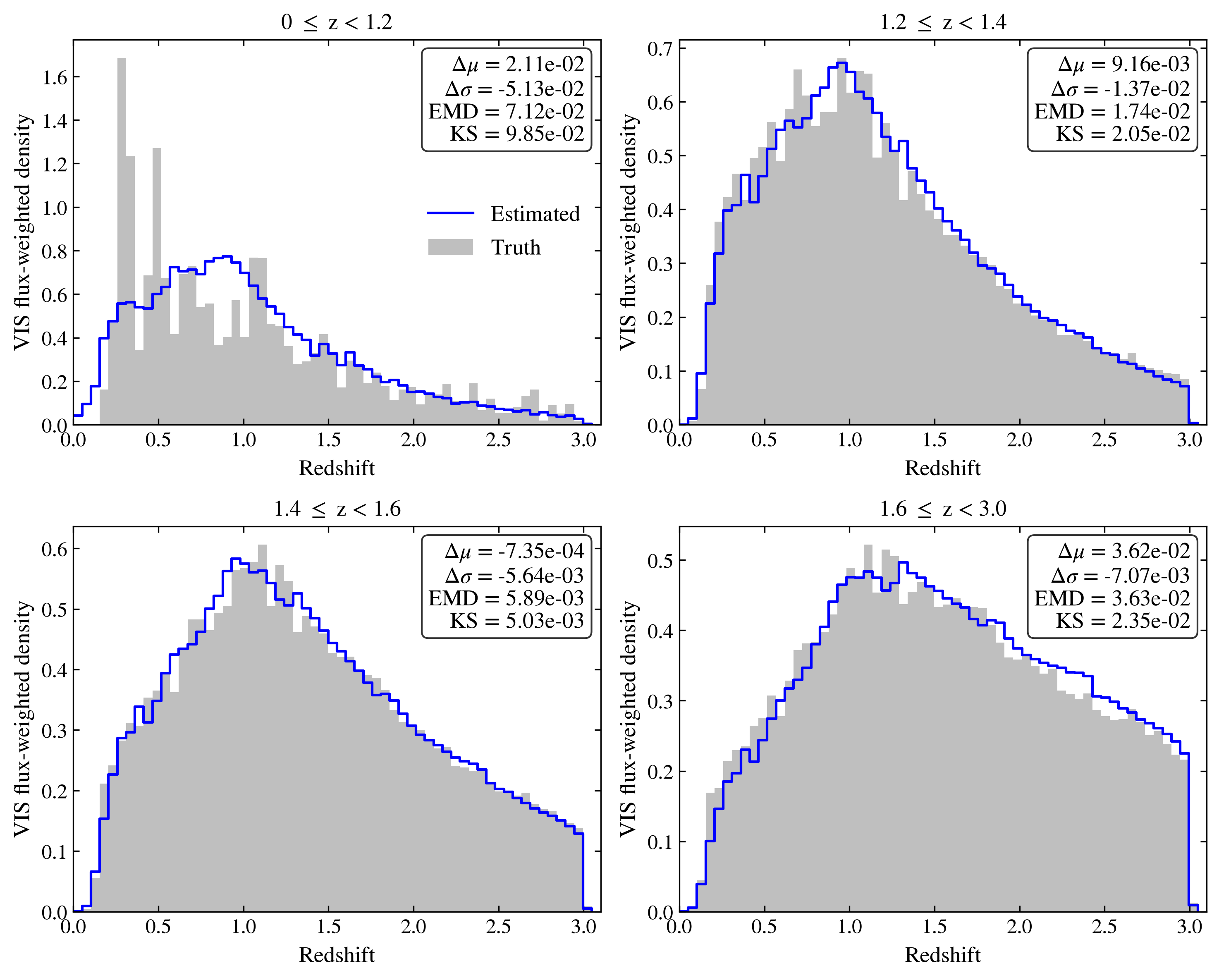}
      \caption{Comparison of VIS flux-weighted redshift distributions between model predictions (lines) and the underlying truth (shaded histograms) for tomographic subsamples defined by pixel-level redshift binning. Each panel corresponds to a different tomographic bin, with the pixel mean redshift range indicated in the panel title.}
         \label{fig:tomo}
  \end{figure*}

The results are shown in Figure~\ref{fig:tomo}. The predicted redshift distributions per tomographic bin exhibit good performance, achieving sub-per cent or near sub-per cent accuracy across all bins, with the largest deviations occurring at low and high redshift corresponding to the edges of the training set. The only noticeable discrepancy appears in the lowest tomographic bin, where the model fails to capture sharp features present in the true distribution. These features typically arise from sample variance within the relatively small cosmological volume probed at low redshifts, which the model inherently smooths over during its statistical learning process.

Although this near sub-per cent accuracy meets the general requirements for tomographic analyses in traditional galaxy-based studies such as weak lensing and photometric galaxy clustering \citep{Ma2006ApJ...636...21M,Euclid2026arXiv260400805E}, further cosmology-focused investigation is warranted to demonstrate the actual constraining power of this pixel-level tomographic approach. Two potential issues are especially worth exploring in future work.

First, the redshift distributions of the tomographic bins overlap substantially. The combination of light from many galaxies within each pixel is expected to contribute to this strong overlap: at our adopted resolution, a pixel contains, on average, ${\sim}130$ undetected galaxies, potentially spanning a broad redshift range, yet receives a single tomographic assignment. Detected-galaxy analyses can assign galaxies individually, although their calibrated redshift distributions also exhibit overlap (e.g. \citealt{Wright2025AA...703A.144W,Yin2025arXiv251023566Y}). The extent of this overlap depends on the sample selection, available photometry, and binning scheme~\citep{Kitching2019PhRvD..99f3536K}. Comparisons must also account for the different weighting conventions, since our distributions are VIS flux-weighted, whereas galaxy-based analyses typically use number or lensing weights. Furthermore, the consequences of overlap depend strongly on the science case, including which observables are used and which cosmological parameters are constrained~\citep{Zuntz2021OJAp....4E..13Z}. Translating a measure of distributional overlap into a cosmological figure of merit therefore requires a dedicated analysis beyond the scope of this work.

Second, our method recovers a flux-weighted redshift distribution, and the number of galaxies contributing to a given pixel is not directly constrained. For map-based analyses this is largely sufficient, since the intensity map is intrinsically a flux-weighted observable and thus requires an intensity-weighted redshift kernel~\citep[cf.][]{Schmidt2015MNRAS.446.2696S,Chiang2019ApJ...877..150C}. It becomes relevant only when the discreteness of the sources must be modelled explicitly, as in the shot-noise term of auto-spectrum analyses \citep{Helgason2012ApJ...752..113H} or of cross-spectra retaining small scales \citep{Cheng2022ApJ...925..136C}. We leave such downstream cosmological analyses to future work.

\section{Conclusions}
\label{Sec:con}

We have presented the first framework for estimating the redshift distributions of undetected galaxies from multi-band maps of unresolved background light using conditional normalising flows. Trained on Flagship-based simulations with \textit{Euclid} and LSST imaging characteristics, the model achieves sub-per cent accuracy in both the mean and standard deviation of the VIS flux-weighted redshift distribution when the training and target samples are statistically well matched. We find that tailored data imputation can largely restore performance under incomplete photometric coverage, whereas mismatches between the training and target samples remain challenging. To diagnose these mismatches, we propose including observable quantities, such as the VIS background noise RMS, among the target variables. Finally, we demonstrate that the model retains sufficient pixel-level information to recover the redshift distributions of tomographically defined subsamples with sub-per cent or near sub-per cent accuracy across all bins.

Several avenues for future development remain. Most notably, potential mismatches in galaxy populations between the training and target samples cannot be fully captured within our current simulation framework and are difficult to diagnose in practice, since the intrinsic properties of real data are not known a priori. In particular, the reported accuracy applies to the galaxy population represented in the input Flagship mock catalogue, which has a finite depth. Assessing the accuracy achievable for real unresolved-background maps therefore requires testing the contribution of galaxies beyond the catalogue's completeness limits. A promising direction is to train multiple models on mock catalogues spanning different depths and modelling assumptions, constructed using hydrodynamical simulations~\citep[e.g.][]{Nelson2019ComAC...6....2N,Schaye2023MNRAS.526.4978S}, semi-analytical models~\citep[e.g.][]{Henriques2015MNRAS.451.2663H,Lacey2016MNRAS.462.3854L,Lagos2018MNRAS.481.3573L}, and empirical prescriptions~\citep[e.g.][]{Carretero2015MNRAS.447..646C,Behroozi2019MNRAS.488.3143B}. The spread in predictions can then provide an estimate of the systematic uncertainty associated with these modelling choices, to be propagated through downstream applications.

It would also be valuable to assess the impact of additional astrophysical backgrounds on model performance and whether the framework can be extended to separate or marginalise over their contributions. Our current simulations include only the undetected galaxy light and instrumental noise, but real observations will also contain diffuse emission from sources such as Galactic cirrus~\citep[e.g.][]{Schlegel1998ApJ...500..525S,Planck2014AA...571A..11P}, intracluster light~\citep[e.g.][]{Montes2022NatAs...6..308M}, and zodiacal light residuals~\citep[e.g.][]{Kelsall1998ApJ...508...44K}. In principle, the flexibility of normalising flows in modelling multi-dimensional target distributions makes them well suited to this task, as additional observable diagnostics -- analogous to the noise RMS employed in this work -- could be introduced to help the model distinguish between the various astrophysical components. Quantifying the impact of these foregrounds -- if they cannot be subtracted a priori, for example with tools like DeNeb (Bertin et al. in prep.) -- on redshift estimation accuracy will be an important step towards applying our framework to real observational data.

Another important avenue is the exploration of finer pixel resolutions. In this work, we adopted $N_{\mathrm{side}} = 4096$, yielding an average of ${\sim}130$ undetected galaxies per pixel, which provides a high signal-to-noise ratio for robust measurement. However, finer resolutions with significantly fewer galaxies per pixel would be desirable for probing smaller-scale structure and would test the limits of the method. This may also require strategies for handling pixels with very few or no galaxies. Determining the optimal map resolution requires balancing the requirements of the specific scientific application against survey sensitivity and remains a valuable direction for future investigation.

From a modelling perspective, a natural extension would be to adopt an ensemble-flow framework~\citep[e.g.][]{Lakshminarayanan2016arXiv161201474L}. Such an approach would not only improve predictive accuracy through model averaging but also provide uncertainty estimates derived from inter-model variance, particularly when the aforementioned systematics are incorporated into the training and model construction.

Overall, our results demonstrate that conditional normalising flows, combined with realistic image simulations, provide a viable and robust approach for estimating the redshifts of undetected galaxy populations from multi-band maps of unresolved background light. The scientific applications of this framework and its extensions are promising. The per-pixel redshift information allows large-scale structure to be probed through spatial correlation analyses of the unresolved background, complementary to equivalent analyses with traditional galaxy surveys~\citep{Heymans2021AA...646A.140H,Sugiyama2023PhRvD.108l3521S,DES2026arXiv260114559D}. Flux-weighted redshift distributions estimated for individual photometric bands further enable cross-correlations among the corresponding band maps to test galaxy population models, complementary to the halo-modelling approach commonly adopted for this purpose~\citep{Planck2014AA...571A..30P,Feder2025ApJ...994...30F}. The approach can be further extended by cross-correlating with detected galaxy catalogues to probe the high-redshift galaxy populations that contribute to the epoch of reionisation~\citep[see e.g.][]{Cheng2022ApJ...925..136C}, or, looking further ahead, applied to aid component separation in future intensity mapping surveys, where the sky background will constrain high-redshift undetected galaxy populations \citep[e.g. the SPHEREx--\textit{Euclid} analysis of][]{Euclid2026arXiv260121111E}.

\begin{acknowledgements}
SSL acknowledges funding from the programme ``Netzwerke 2021'', an initiative of the Ministry of Culture and Science of the State of North Rhine-Westphalia. LVW is supported by the Natural Sciences and Engineering Research Council of Canada. Some of the calculations for this publication were performed using the HPC cluster Elysium of the Ruhr University Bochum, subsidised by the DFG (INST 213/1055-1), and the resources of the SLAC Shared Science Data Facility (S3DF) at SLAC National Accelerator Laboratory. This work has made use of CosmoHub, developed by PIC (maintained by IFAE and CIEMAT) in collaboration with ICE-CSIC. It received funding from the Spanish government (grant EQC2021-007479-P funded by MCIN/AEI/10.13039/501100011033), the EU NextGeneration/PRTR (PRTR-C17.I1), and the Generalitat de Catalunya.

\end{acknowledgements}

  {\footnotesize
  \bibliographystyle{aa}
  \bibliography{reference}}

\begin{appendix}

\section{Conditional normalising flows}
\label{Sec:flow}

This appendix provides an overview of the essential concepts underlying conditional normalising flows, focusing on the architectures adopted in our model. For comprehensive introductions, we refer the reader to recent reviews by \citet{Kobyzev2019arXiv190809257K} and \citet{Papamakarios2019arXiv191202762P}.

A normalising flow is a generative model capable of learning complex probability distributions whilst allowing tractable density evaluation. It employs a sequence of parameterised, invertible, and differentiable transformations — known as bijections — to map a simple base probability distribution, such as a standard normal or uniform distribution, into a complex target distribution. Let $\bm{u} \sim p_{\mathrm{base}}(\bm{u})$ be a sample from the base distribution and $f$ the bijection. Then $\bm{x} = f(\bm{u})$ follows a new distribution whose density is given by the change-of-variables formula:
\begin{equation}
\label{eq:change_of_variables}
p(\bm{x}) = p_{\mathrm{base}}\!\left(f^{-1}(\bm{x})\right) \left|\det \frac{\partial f^{-1}}{\partial \bm{x}}\right|.
\end{equation}
If $f$ is sufficiently expressive, it can in principle map any base distribution to any target distribution \citep[e.g.][]{Bogachev2005SbMat.196..309B}.

The term `flow' reflects the fact that $f$ can be constructed as a composition of simple invertible layers, $f = f_L \circ \cdots \circ f_{\ell} \circ \cdots \circ f_1$. With the intermediate representations defined as $\bm{y}^{(0)} = \bm{u}$ and $\bm{y}^{(\ell)} = f_\ell(\bm{y}^{(\ell-1)})$, so that $\bm{x} = \bm{y}^{(L)}$, each layer incrementally reshapes the distribution. Because the composition of invertible functions is itself invertible, and the Jacobian determinant of the composition factorises as
\begin{equation}
\left|\det \frac{\partial f^{-1}}{\partial \bm{x}}\right| = \prod_{\ell=1}^{L} \left|\det \frac{\partial f_\ell^{-1}}{\partial \bm{y}^{(\ell)}}\right|,
\end{equation}
both sampling and density evaluation remain tractable.

The primary architectural challenge lies in designing $f$ such that it is expressive yet computationally efficient. One of the most widely adopted solutions is the coupling layer~\citep{Dinh2014arXiv1410.8516D}, which partitions the input $\bm{y}^{(\ell -1)}_{1:D}$ into two groups: $\bm{y}^{(\ell -1)}_{1:d}$ and $\bm{y}^{(\ell -1)}_{d+1:D}$. The first group is passed through unchanged, $\bm{y}^{(\ell)}_{1:d} = \bm{y}^{(\ell -1)}_{1:d}$, whilst the second is transformed element-wise by an invertible function (the transformer), whose parameters are determined by the first group through a neural network (NN) known as the conditioner:
\begin{equation}
\bm{y}^{(\ell)}_{d+1:D} = g_{\bm{\theta}}(\bm{y}^{(\ell -1)}_{d+1:D})\,, \quad \text{where} \quad \bm{\theta} = \mathrm{NN}(\bm{y}^{(\ell-1)}_{1:d}).
\end{equation}

Under this construction, the Jacobian of each layer $\ell$ is a block lower-triangular matrix whose determinant is efficiently computable as $\prod_{i=d+1}^{D} \partial g_{\theta_i}/\partial y^{(\ell-1)}_i$. To ensure that all dimensions are eventually transformed, multiple coupling layers are stacked with random permutations of the dimensions between successive layers~\citep{Dinh2014arXiv1410.8516D,Dinh2016arXiv160508803D,Kingma2018arXiv180703039K}.

The choice of element-wise transformer $g_{\bm{\theta}}$ is flexible, provided it is differentiable and invertible. Whilst early models relied on simple additive or affine transformations, a highly expressive modern alternative is the monotonic rational-quadratic spline \citep{Durkan2019arXiv190604032D}. This approach divides a specified bounding interval $[-B, B]$ into $K$ discrete segments delineated by $K+1$ knots. Within each segment, the transformation is governed by a rational-quadratic function, which is the quotient of two quadratic polynomials. The NN conditioner outputs unconstrained values that are passed through activation functions, such as softmax and softplus, to yield strictly positive bin widths, bin heights, and knot derivatives. This construction ensures strict monotonicity, guaranteeing analytical invertibility via the standard quadratic formula. For inputs falling outside $[-B, B]$, the transformation defaults smoothly to linear tails for numerical stability.

\begin{figure*}
\centering
\begin{adjustbox}{max width=\textwidth}
\begin{tikzpicture}[
    >=stealth,
    node distance=1.2cm,
    box/.style={rectangle, draw=green!60!black, fill=green!8, rounded corners=3pt,
                minimum height=0.9cm, minimum width=2.8cm, font=\small, align=center},
    passbox/.style={box, minimum width=2.6cm},
    transbox/.style={box, minimum width=2.6cm},
    distbase/.style={rectangle, draw=green!60!black, fill=green!8, rounded corners=8pt,
                     minimum height=2.2cm, minimum width=1.8cm, font=\small, align=center},
    disttarget/.style={rectangle, draw=green!60!black, fill=green!25, rounded corners=8pt,
                       minimum height=2.2cm, minimum width=1.8cm, font=\small, align=center},
    condbox/.style={rectangle, draw=blue!40, fill=blue!8, rounded corners=3pt,
                    minimum height=0.9cm, minimum width=3.6cm, font=\small, align=center},
    permute/.style={circle, draw=gray!60, fill=gray!15, minimum size=0.55cm, font=\tiny},
    arrstyle/.style={->, thick, draw=green!50!black},
    condarr/.style={->, thick, draw=blue!50!black},
    every node/.style={font=\small}
]

\node[distbase] (base) at (0, 0) {Base\\[2pt]$\bm{u} \sim p_{\mathrm{base}}(\bm{u})$};

\node[passbox] (L1top) at (3.8, 1.0) {$\bm{y}_{1:d}^{(1)} = \bm{u}_{1:d}$};
\node[transbox] (L1bot) at (3.8, -1.0) {$\bm{y}_{d\!+\!1:D}^{(1)} = g_{\bm{\theta}}(\bm{u}_{d\!+\!1:D})$};

\draw[arrstyle] (base.east) ++(0, 0.3) -- ++(0.4, 0) |- (L1top.west);
\draw[arrstyle] (base.east) ++(0,-0.3) -- ++(0.4, 0) |- (L1bot.west);

\draw[arrstyle] (L1top.south) -- (L1bot.north);

\node[permute] (P1) at (6.2, 0) {$\bm{\pi}$};
\draw[arrstyle] (L1top.east) -| (P1.north);
\draw[arrstyle] (L1bot.east) -| (P1.south);

\node[passbox] (Lltop) at (8.6, 1.0) {$\bm{y}_{1:d}^{(\ell)} = \bm{y}_{1:d}^{(\ell-1)}$};
\node[transbox] (Llbot) at (8.6, -1.0) {$\bm{y}_{d\!+\!1:D}^{(\ell)} = g_{\bm{\theta}}(\bm{y}_{d\!+\!1:D}^{(\ell-1)})$};

\draw[arrstyle] (P1.north) |- (Lltop.west);
\draw[arrstyle] (P1.south) |- (Llbot.west);

\draw[arrstyle] (Lltop.south) -- (Llbot.north);

\node[permute] (P2) at (11.0, 0) {$\bm{\pi}$};
\draw[arrstyle] (Lltop.east) -| (P2.north);
\draw[arrstyle] (Llbot.east) -| (P2.south);

\node[passbox] (LLtop) at (13.4, 1.0) {$\bm{y}_{1:d}^{(L)} = \bm{y}_{1:d}^{(L\!-\!1)}$};
\node[transbox] (LLbot) at (13.4, -1.0) {$\bm{y}_{d\!+\!1:D}^{(L)} = g_{\bm{\theta}}(\bm{y}_{d\!+\!1:D}^{(L\!-\!1)})$};

\draw[arrstyle] (P2.north) |- (LLtop.west);
\draw[arrstyle] (P2.south) |- (LLbot.west);

\draw[arrstyle] (LLtop.south) -- (LLbot.north);

\node[disttarget] (target) at (17.2, 0) {Target\\[2pt]$\bm{x} \sim p(\bm{x}\,|\,\bm{c})$};

\draw[arrstyle] (LLtop.east) ++(0,0) -- ++(0.4, 0) |- ([yshift=3pt]target.west);
\draw[arrstyle] (LLbot.east) ++(0,0) -- ++(0.4, 0) |- ([yshift=-3pt]target.west);

\node[condbox] (cond) at (8.6, -3.8) {Conditioning variables~$\bm{c}$};

\draw[condarr] (cond.north) ++(0,0) -- (Llbot.south)
    node[midway, right=2pt, font=\footnotesize] {$\bm{\theta} = \mathrm{NN}(\bm{y}_{1:d}^{(\ell-1)},\, \bm{c})$};

\draw[condarr] (cond.north west) ++(0.3,0) -- (L1bot.south)
    node[midway, left=2pt, font=\footnotesize] {$\bm{\theta} = \mathrm{NN}(\bm{u}_{1:d},\, \bm{c})$};

\draw[condarr] (cond.north east) ++(-0.3,0) -- (LLbot.south)
    node[midway, right=2pt, font=\footnotesize] {$\bm{\theta} = \mathrm{NN}(\bm{y}_{1:d}^{(L-1)},\, \bm{c})$};

\node[font=\footnotesize, gray] at (3.8, 2.0) {Coupling layer 1};
\node[font=\footnotesize, gray] at (8.6, 2.0) {Coupling layer $\ell$};
\node[font=\footnotesize, gray] at (13.4, 2.0) {Coupling layer $L$};

\end{tikzpicture}
\end{adjustbox}
\caption{Schematic overview of the conditional normalising flow architecture with coupling layers. The base distribution $\bm{u} \sim p_{\mathrm{base}}(\bm{u})$ is progressively transformed through $L$ coupling layers into the target distribution $p(\bm{x}\,|\,\bm{c})$. Within each coupling layer, the input is partitioned into two groups: the first $d$ dimensions pass through unchanged, whilst the remaining dimensions are transformed element-wise by the transformer $g_{\bm{\theta}}$. The transformer parameters $\bm{\theta}$ are derived from the unchanged $d$ dimensions of the input together with the conditioning variables $\bm{c}$ through the neural network conditioner. Random permutations $\bm{\pi}$ between successive layers ensure that all dimensions are eventually transformed. In our application, the base distribution is a multi-dimensional standard normal distribution whose dimensionality matches that of the target distribution, which comprises the desired properties of the undetected galaxy population, including the redshift and magnitude distributions, along with any observable quantities included for model validation. The conditioning variables $\bm{c}$ correspond to the features derived from the ten-band HEALPix map pixel values of unresolved background light and are fed into every conditioner network, allowing the flow to learn a distinct transformation for each set of observed pixel values, thereby enabling the estimation of underlying undetected galaxy properties from observed multi-band background light.}
\label{fig:flow_schematic}
\end{figure*}
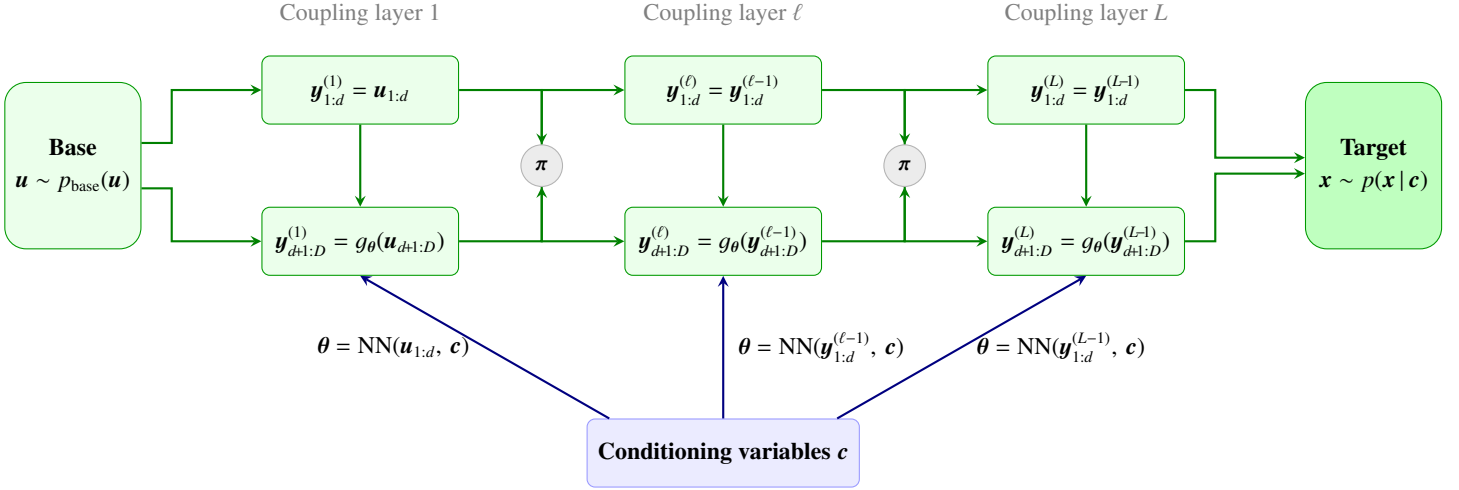

Building upon this foundational flow architecture, one can introduce external conditioning variables, $\bm{c}$, to learn conditional probability distributions, $p(\bm{x}\,|\,\bm{c})$. The resulting model is known as a conditional normalising flow. In practice, this is achieved by incorporating the conditioning variables as additional inputs to the NN conditioner, such that $\bm{\theta} = \text{NN}(\bm{y}^{(\ell-1)}_{1:d}, \bm{c})$. By explicitly conditioning the bijections on this additional information, the flow learns a distinct transformation for every possible value of $\bm{c}$, thereby enabling conditional data generation and density evaluation. A schematic overview of this architecture is shown in Figure~\ref{fig:flow_schematic}.

In our study, we draw $\bm{u}$ from a three-dimensional standard normal distribution, while the transformed variable $\bm{x} = f(\bm{u})$ represents the three targeted quantities of the undetected galaxy population, namely its redshift, VIS magnitude, and the RMS of the VIS imaging noise. The intermediate representations $\bm{y}^{(\ell -1)}_{1:D}$ share this dimensionality, so $D=3$. The partition size is then $d=\lfloor D/2 \rfloor=1$: one dimension is left untransformed at each layer, serving (together with the conditioning variables $\bm{c}$) as input to the conditioner network, while the remaining $D-d=2$ dimensions are transformed. The conditioning variables $\bm{c}$ refer to the features derived from the pixel values of the ten-band HEALPix maps. The flow consists of $L=8$ coupling layers, each followed by a random permutation of the dimensions, ensuring that all three dimensions are sufficiently transformed throughout the flow. The bounding interval is $B=5$, and the number of spline knots is $K+1=16$. The conditioner is implemented as a multi-layer perceptron with four hidden layers, each of width 256, using rectified linear unit activation functions. Further training and validation details are provided in Section~\ref{Sec:training}.

\end{appendix}

\end{document}